\documentclass{article}
\pdfoutput=1

\usepackage{arxiv}
\usepackage[export]{adjustbox}
\usepackage[para,online,flushleft]{threeparttable}
\usepackage{booktabs}
\usepackage{threeparttable}
\usepackage{makecell}
\usepackage[utf8]{inputenc} 
\usepackage[T1]{fontenc}    
\usepackage{hyperref}       
\usepackage{url}            
\usepackage{booktabs}       
\usepackage{amsfonts}       
\usepackage{nicefrac}       
\usepackage{microtype}      

\usepackage{graphicx}
\usepackage[caption=false]{subfig}
\usepackage{doi}
\usepackage{amsmath}
\usepackage{multirow}
\usepackage{multicol}
\usepackage{titlesec}
\titlespacing*{\section}{0pt}{0.3\baselineskip}{0.3\baselineskip}
\titlespacing*{\paragraph}{0pt}{0.3\baselineskip}{0.3\baselineskip}

\title{Uncertainty-Aware Complex Scientific Table Data Extraction}

\author{Kehinde Ajayi\\
	Old Dominion University\\
	Norfolk, VA, 23529\\
	\texttt{kajay001@odu.edu} \\
    \And
    Yi He\\
    William \& Mary\\
    Williamsburg, VA, 23187 \\
    \texttt{yihe@wm.edu} \\
   \And
    Jian Wu\\
	Old Dominion University\\
	Norfolk, VA, 23529 \\
	\texttt{j1wu@odu.edu} \\
}

\renewcommand{\shorttitle}{Uncertainty-Aware Complex Scientific Table Data Extraction}

\begin{document}
\maketitle

\begin{abstract}
Table structure recognition (TSR) and optical character recognition (OCR) play crucial roles in extracting structured data from tables in scientific documents. However, existing extraction frameworks built on top of TSR and OCR methods often fail to quantify the uncertainties of extracted results. To obtain highly accurate data for scientific domains, all extracted data must be manually verified, which can be time-consuming and labor-intensive. We propose a framework that performs uncertainty-aware data extraction for complex scientific tables, built on conformal prediction, a model-agnostic method for uncertainty quantification (UQ).  We explored various uncertainty scoring methods to aggregate the uncertainties introduced by TSR and OCR. We rigorously evaluated the framework using a standard benchmark and an in-house dataset consisting of complex scientific tables in six scientific domains. The results demonstrate the effectiveness of using UQ for extraction error detection, and by manually verifying only 47\% of extraction results, the data quality can be improved by 30\%. Our work quantitatively demonstrates the role of UQ with the potential of improving the efficiency in the human-machine cooperation process to obtain scientifically usable data from complex tables in scientific documents. All code and data are available on GitHub at \url{https://github.com/lamps-lab/TSR-OCR-UQ/tree/main}.
\end{abstract}


\section{Introduction}
Tables are a fundamental data representation format used extensively in scientific literature, financial reports, and structured documents \cite{yildiz2005pdf2table,chen2013automatic,pinto2003table}. Extracting data from table images is crucial for data analysis, enabling downstream applications such as knowledge extraction \cite{rastogi2020information}, information retrieval \cite{mitra2000information}, and automated data processing \cite{holub2021toward}. This process generally involves Table Structure Recognition (TSR) to identify the spatial arrangement of rows, columns, and cells \cite{deeptabstr,cascadetabnet,li2021rethinking}, and Optical Character Recognition (OCR) to extract textual content. 

\begin{figure}
    \centering
    \includegraphics[width=1.0\textwidth]{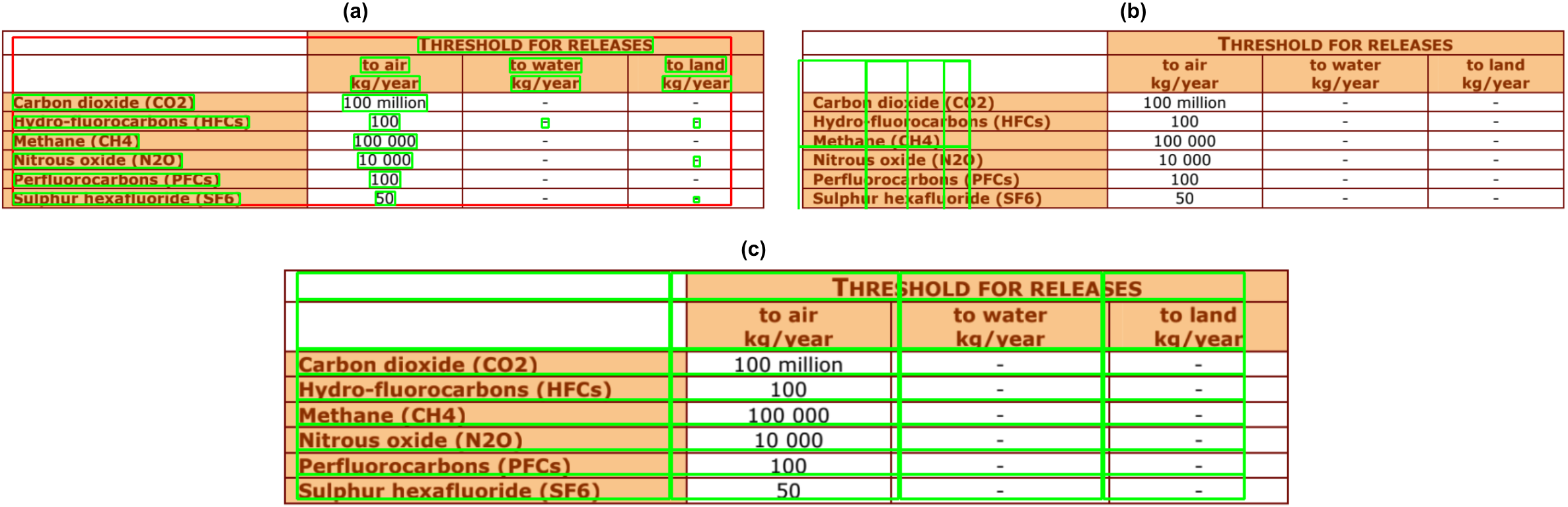}
     \caption{(a) and (b) represent the results of OCR and TSR respectively, and (c) represents the results after they are properly integrated.}
     \label{fig:tsr-ocr}
\end{figure}

Although recent advances in deep learning have significantly improved individual components of table extraction, integrating TSR and OCR into a unified pipeline remains a major challenge due to structural misalignment, OCR inaccuracies, and the lack of uncertainty estimation in extracted data. TSR methods focus on detecting table cell boundaries, whereas OCR is applied either independently or directly on TSR-detected structures to extract text. This often results in discrepancies where text is incorrectly assigned to table cells, misaligned due to imperfect bounding box detection, or even omitted when OCR struggles with complex table layouts (see Fig.~\ref{fig:tsr-ocr}). This issue becomes even more pronounced in the context of complex scientific table data extraction because of a wide range of structural diversity, including spanning cells, multi-row headers, and nested tables, which further complicates alignment. Extraction accuracy can be further degraded by OCR errors resulting from poor image quality or domain-specific content.

Little work has been done to integrate TSR and OCR outputs (TSR-OCR) e.g., \cite{wang2021tablelab}, and most existing work lacks a mechanism to quantify uncertainty in extracted data, making it challenging to identify and correct errors before they propagate into downstream applications.

Uncertainty quantification (UQ) is a critical aspect of machine learning and statistical modeling, aiming to assess the confidence of model predictions. Among various UQ methods such as Bayesian models \cite{park2010bayesian} and Monte Carlo dropout \cite{gal2016dropout}, Conformal Prediction (CP) stands out because it provides statistically valid prediction intervals without making strong assumptions about the underlying model or data distribution \cite{angelopoulos2021gentle}. This technique constructs prediction sets that, with a predefined confidence level, are guaranteed to contain the true outcome, thereby offering a measure of reliability for each prediction. Despite its successful application in medical diagnostics \cite{kutiel2023conformal,vazquez2022conformal} and earth sciences \cite{singh2024uncertainty}, its integration with table data extraction remains under-explored. The inherent uncertainties in TSR and OCR outputs—stemming from factors like complex table layouts and varying image qualities—necessitate a robust UQ approach. We chose CP for our approach because of its model-agnostic nature \cite{ye2024benchmarking} and because it does not require probabilistic model assumptions or additional training, making it computationally efficient \cite{angelopoulos2021gentle} and well-suited to enhance the reliability of integrated TSR-OCR pipelines.

In this study, we propose the TSR-OCR-UQ framework, which leverages UQ to improve data extraction quality from complex scientific tables. We examine whether confidence scores derived from OCR alone are sufficient to detect extraction errors or whether a more robust CP-based uncertainty measure is necessary. Furthermore, we explore how well various conformal scoring methods impact the performance of UQ in terms of identifying incorrect extractions and how integrating a UQ-informed human verification mechanism, including comparing automatically extracted data and ground truth data and correcting erroneous results, improves overall data quality and efficiency. Our approach integrates TSR-predicted cell structures with OCR text extraction and applies CP-based UQ to estimate the uncertainty of extracted data. The framework automatically flags cells with high uncertainty. We evaluate our framework on GenTSR \cite{ajayi2023study}, a dataset consisting of diverse scientific tables from multiple disciplines, and the ICDAR 2013 \cite{icdar2013competition} dataset, a standard benchmark for TSR. 

In our study, we employ PaddleOCR \cite{li2022pp} as the OCR engine. Our experiments that compare state-of-the-art OCR engines, including PaddleOCR \cite{li2022pp}, EasyOCR \cite{jaided2020easyocr}, and docTR \cite{doctr2021}, using table images from three academic domains consistently demonstrated that PaddleOCR outperformed the others. For TSR, we adopt Table Transformer (TATR) \cite{smock2021tabletransformer}, a deep learning-based model known for its effectiveness in accurately predicting table structures. It is worth noting that our approach is independent of the specific OCR and TSR models employed. The only requirement for integration into our UQ framework is that the models output confidence scores for their predictions. This requirement can be easily met by most deep learning architectures, where confidence scores are naturally provided as softmax outputs. As a result, our framework can be customized to adopt other TSR and OCR models.

Our contributions are as follows:
\begin{itemize}
    \item We propose an integrated framework that automatically performs uncertainty-aware table data extraction.

    \item We demonstrate that integrating TSR and OCR outputs by constructing grid cells enables alignment between textual content and structural elements of tables.
    
    \item We demonstrate that conformal prediction-based UQ effectively quantifies uncertainty in extracted data and enables automated error flagging, improving the quality of table data extraction.  

    \item We compared various conformal scoring methods that quantify the uncertainties introduced by TSR and OCR, highlighting the most reliable approaches for uncertainty-aware table data extraction.

    \item We demonstrate that using our framework, UQ can improve the data quality by approximately 30\% and  reduce data verification effort by 53\%.
\end{itemize}

\section{Related Work}
\subsection{Table Detection and Table Structure Recognition}
In the domain of table detection and table structure recognition, early methodologies predominantly relied on heuristic-based approaches \cite{babatunde2015automatic}. For instance, systems such as pdf2table \cite{yildiz2005pdf2table} utilized structural rules and metadata to parse tables from PDF documents, converting textual elements into structured formats such as XML. Although these methods are effective in specific scenarios, they often lack generalizability, particularly when confronted with complex table structures prevalent in scientific literature.

To address these limitations, probabilistic models such as Conditional Random Fields (CRFs) and Hidden Markov Models were introduced. Pinto et al. \cite{pinto2003table} employed CRFs to segment and classify table components, including headers and data cells, by analyzing their spatial and structural relationships. This approach demonstrated improved accuracy over heuristic methods, yet challenges persisted in adapting to the diverse and intricate structures of complex scientific tables.

Recent advancements in deep learning have led to end-to-end models capable of detecting table regions within documents and identifying their internal row and column structures \cite{cascadetabnet,Graph-based-TSR-lee,Multi-Type-TD-TSR,TabStructNet,TGRNet,paliwal2019tablenet}. Despite these advancements, these models primarily focus on recognizing structural boundaries and usually do not perform content extraction—the step that captures the actual data or text within each table cell. This limitation underscores the necessity for integrated approaches that combine table structure recognition with content extraction to produce fully structured data for downstream tasks.

In the context of table data extraction, integrating OCR systems has been explored to enhance content retrieval from scanned images or low-quality PDFs \cite{wang2021tablelab}. Domain-specific frameworks have been developed to manage hierarchical structures and implicit relationships between headers and cells. For example, Chen and Cafarella \cite{chen2013automatic} investigated the extraction of relational data from web spreadsheets, highlighting the complexities introduced by human-designed structures and non-standard schemas. Their system employed automated relational extraction processes, yet necessitated user intervention for fine-tuning, emphasizing the challenges inherent in domain-specific adaptations.

Although the integration of deep learning-based methods with OCR tools has improved content extraction \cite{wang2021tablelab}, the incorporation of UQ within this context remains under-explored. 

\subsection{Uncertainty Quantification}
Uncertainty quantification is a critical aspect of machine learning, providing insights into the reliability of model predictions. Mainstream UQ methods include Bayesian models \cite{park2010bayesian}, Monte Carlo Dropout \cite{gal2016dropout}, and ensemble techniques \cite{lakshminarayanan2017simple}. However, these approaches often involve significant computational complexity and may rely on specific assumptions about data distributions.

CP \cite{shafer2008tutorial} has emerged as a promising UQ method, offering statistically valid prediction sets without stringent data distribution assumptions \cite{ye2024benchmarking}. CP operates by assigning uncertainty estimates to predictions, thereby enabling the measurement of uncertainty in a model-agnostic manner \cite{angelopoulos2021gentle}. This characteristic makes CP particularly advantageous in applications requiring reliable uncertainty estimates, such as table data extraction.

The application of CP in various domains has demonstrated its efficacy in providing robust uncertainty measures. For instance, in Earth Observation data, CP has been utilized to generate prediction intervals that account for inherent uncertainties, thereby enhancing the reliability of datasets used in environmental monitoring \cite{singh2024uncertainty}. Similarly, in medical imaging, CP has been applied to quantify uncertainties in segmentation tasks, offering statistically valid confidence intervals that are crucial for clinical decision-making \cite{gade2024impact}.

In the realm of table data extraction, the integration of CP can quantify the uncertainties arising from OCR inaccuracies and recognition of complex table structures. By providing prediction sets with associated uncertainty estimates, CP enables practitioners to assess the trustworthiness of extracted data, facilitating more informed and reliable analyses. This approach not only enhances the robustness of data extraction pipelines but also contributes to the development of uncertainty-aware systems capable of handling the complexities inherent in complex scientific tables.

\begin{figure*}
    \centering
    \includegraphics[width=1.0\textwidth]{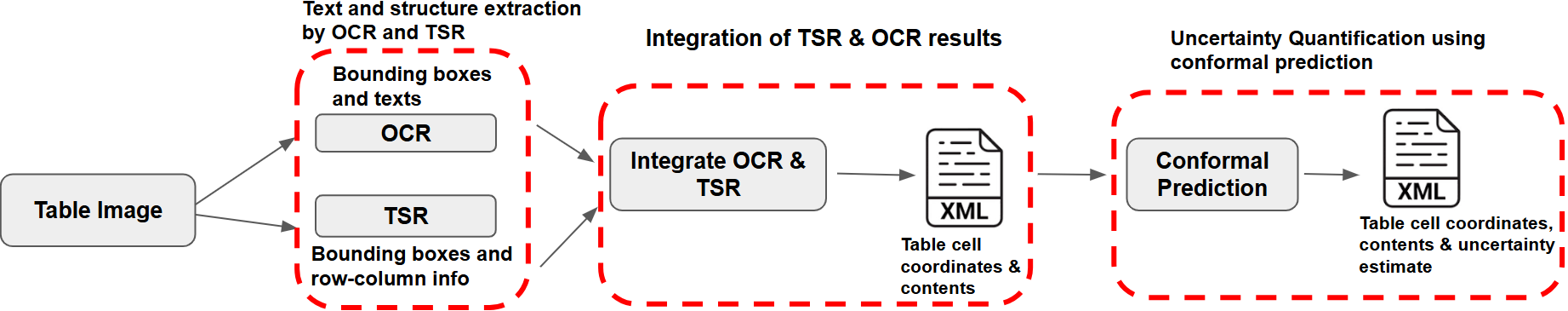}
     \caption{A schematic illustration of the proposed UQ pipeline (\texttt{TSR-OCR-UQ}).}
     \label{fig:uq_pipeline}
\end{figure*}

\section{TSR-OCR-UQ Framework}
Figure~\ref{fig:uq_pipeline} illustrates the architecture of our framework that integrates TSR, OCR, and UQ. The key modules include (1) Text extraction by PaddleOCR, and TSR by TATR, (2) Integration of TSR and OCR results, and (3) UQ via conformal prediction. 

\subsection{Text Extraction and Structure Recognition}
To ensure high-quality text extraction in our TSR-OCR-UQ framework, we conducted an empirical evaluation of three OCR engines—PaddleOCR \cite{du2020pp}, docTR \cite{doctr2021}, and EasyOCR \cite{jaided2020easyocr}—on the GenTSR dataset \cite{ajayi2023study}. This dataset consists of 386 table images cropped from research papers in six domains: Computer Science, Materials Science, Chemistry, Business, Economics, and Biology, providing a diverse dataset for assessing OCR performance in scientific tables. The ground truth annotations include manually labeled bboxes for table cells, along with structural properties such as start-row, start-column, end-row, and end-column. These tables are complex because of their structural variability, including irregular layouts, multi-row/multi-column spanning cells, nested tables, and dense mathematical symbols compared to standard benchmarks such as ICDAR 2013 \cite{icdar2013competition} and ICDAR 2019 \cite{gao2019icdartable}.

\subsubsection{OCR Selection and Comparison}
Each OCR engine differs in its underlying architecture and approach to text recognition:

\begin{itemize}
    \item {\bf PaddleOCR} \cite{du2020pp}: A lightweight, high-performance OCR engine that employs a differentiable binarization-based detection module and a CRNN-based text recognition model \cite{fu2017crnn}. It supports multilingual text recognition, text rotation correction, and structured text processing, making it particularly well-suited for extracting textual content from complex table structures.

    \item {\bf docTR OCR} \cite{doctr2021}: A deep-learning-based OCR engine designed for document parsing, leveraging a Transformer-based text detection model and CTC-based recognition. While effective in general document analysis, it struggles with recognizing mathematical symbols, leading to slight performance degradation in table-based extraction tasks.

    \item {\bf EasyOCR} \cite{jaided2020easyocr}: An open-source OCR engine that employs a CNN-based detection model \cite{fu2017crnn} with LSTM-based recognition. While it is a lightweight alternative, it struggles with domain-specific content such as notations \cite{salehudin2023analysis}, resulting in lower accuracy when applied to scientific tables.
\end{itemize}

To compare OCR performance, we employed the Levenshtein accuracy metric, which measures the similarity between the extracted text and the ground truth for each table cell by computing the Levenshtein distance—the minimum number of character insertions, deletions, or substitutions required to transform one string into another. The Levenshtein accuracy for each cell is formally defined as:

\begin{equation} 
\text{Levenshtein Accuracy} = 1 - \frac{\text{Levenshtein Distance}(\text{extracted}, \text{ground truth})}{\max(\text{length of extracted}, \text{length of ground truth})} 
\end{equation}

where a score of 1.0 indicates a perfect match between the extracted and ground truth text. Lower values indicate increasing textual discrepancies. We average the Levenshtein accuracy for all cells as the overall accuracy for the whole table.

Table~\ref{tab:ocr_comparison} shows that PaddleOCR consistently achieves the highest accuracy across tables in three scientific domains, making it the preferred choice for our TSR-OCR-UQ framework. While docTR OCR performs competitively, it falls slightly short in structured table extraction tasks. EasyOCR exhibits the weakest performance, particularly struggling with misaligned and multi-line text extractions.

\begin{table}[h]
    \centering
    \caption{Performance comparison of 3 OCR engines.}
    \label{tab:ocr_comparison}
    \begin{tabular}{l|c|c|c|c}
        \toprule
        OCR Engine  & Computer Science & Materials Science & Biology & Average \\
        \midrule
        {\bf PaddleOCR} & {\bf 0.642} & {\bf 0.723} & {\bf 0.779} & {\bf 0.715} \\
         \midrule
        docTR     & 0.642 & 0.701 & 0.769 & 0.704 \\
         \midrule
        EasyOCR   & 0.608 & 0.645 & 0.685 & 0.646 \\
        \bottomrule
    \end{tabular}
\end{table}

For structure recognition, we adopt TATR \cite{smock2021tabletransformer}, a transformer-based object detection model that excels in recognizing table structures such as rows, columns, and cells. We chose TATR based on its performance on established TSR datasets. For instance, TATR achieves an exact match accuracy of 75\% on the ICDAR-2013 benchmark when trained on the PubTables-1M dataset \cite{smock2022pubtables}, surpassing traditional TSR models. Furthermore, TATR attains a 65\% accuracy on the same benchmark (ICDAR-2103) when trained on FinTabNet \cite{zheng2021global} and an 81\% accuracy when trained on a combined dataset, demonstrating its robustness across diverse data sources \cite{smock2023aligning}. These results underscore TATR's superior capability in accurately detecting and structuring complex table components, justifying its adoption in our pipeline.

\subsection{Integration of TSR and OCR Outputs}  
\subsubsection{Grid Cell Construction}  
TSR and OCR provide different predictions. TSR models, such as TATR, predict table structures by detecting rows and columns in a table image, predicting class labels (e.g., 1 for row, 2 for column), and bounding box (bbox) coordinates for {\bf table cells} (e.g., \( (x_0, y_0, x_1, y_1) \)). In contrast, OCR engines recognize text at the character or word level, providing bbox coordinates for individual {\bf text segments} and the recognized characters without grouping them into table cells. This disparity creates misalignment issues, as TSR and OCR operate at different granularities, making it challenging to associate extracted text with its corresponding structural elements.

To address this challenge, we construct grid cells that serve as an intermediate representation to integrate TSR and OCR outputs. We generate a structured grid of table cells by intersecting the row and column bboxes predicted by TSR. Each grid cell represents a single table cell, defined by the intersection of a row bbox and a column bbox from TSR. To preserve structural consistency, we assign each cell a row index, a column index, and confidence scores derived from TSR predictions.

Next, to perform spatial alignment of TSR and OCR results, we scale the bboxes predicted by TSR to match the dimensions of OCR-predicted bboxes, effectively reducing inconsistencies between the predictions of two distinct methods. Additionally, since OCR often fragments text across multiple bboxes, we collapse the bboxes to merge OCR-predicted segments within the same cell, aggregating their coordinates into a single unified bbox. This is particularly important for multi-row spanning cells, ensuring that text is correctly assigned to the appropriate table structure.

\subsubsection{Location Confidence Extraction}  
We assume that both the TSR and OCR models provide confidence scores for their predictions. These confidence scores are typically derived from the softmax layer of deep learning models, including open-weight large language models (LLMs) \cite{liu2024uncertainty}, to indicate the reliability of each prediction.

The TATR model outputs confidence scores for each detected row and column. To estimate the location confidence score for each table cell, we propagate these scores to the corresponding grid cells based on the assigned row and column indices. Specifically, each cell inherits the confidence scores of its associated row and column. We then compute a single location confidence score per cell by taking the average of its assigned row and column confidence scores. This aggregated confidence score serves as a measure of the structural certainty of the TSR model, providing an indicator of how confidently the model recognizes table boundaries and cell positions.

\subsubsection{Matching Texts and Confidence Scores to Grid Cells}
To integrate the outputs of OCR with the grid cells, we match the textual content and confidence scores from OCR to the corresponding table cells generated by the TATR model. We achieve this through a geometric alignment process that leverages spatial overlaps between OCR bboxes and TSR grid cells. Below are the key steps of the process:

\begin{itemize}
    \item {\bf Representation of Geometries}: We convert the OCR-predicted bboxes into polygons using their coordinates, allowing for precise geometric comparisons with the grid cells. Similarly, we represent each grid cell as a polygon to facilitate intersection-based matching.

    \item {\bf Matching OCR Outputs to Grid Cells}: For each OCR bbox, we compute the intersection area (IoA) with a TSR grid cell. If the IoA exceeds 50\% of the OCR bbox’s area, we consider the OCR bbox as a match for that grid cell. This threshold ensures robust alignment for the majority of misaligned bbox and grid cells.
    \item {\bf Aggregation of Text and Confidence Scores}:
    For each matched grid cell, we concatenate the texts from all overlapping OCR bboxes to form the cell’s complete content. Then, we average the confidence scores of the matched OCR bboxes to compute an overall OCR confidence score for the grid cell. 

    \item {\bf Handling Cells with No OCR Match}:  
    In cases where no OCR text is matched to a grid cell (e.g., blank cells with no OCR confidence), we set the OCR confidence of the cell to 0, and retain the structural information predicted by the TSR model and use it for downstream UQ. This ensures that cells without direct OCR alignment can still contribute to the final uncertainty-aware extraction process.
\end{itemize}

\subsection{Uncertainty Quantification using Conformal Prediction}

Conformal prediction (CP) is a statistically model-agnostic framework for uncertainty quantification \cite{ye2024benchmarking}. Given an input instance \( X_t \), CP constructs a prediction set \( C(X_t) \subset \mathcal{Y} \) such that the true label \( Y_t \) is contained within this set with at least a user-specified probability \( 1 - \alpha \). Formally, this is expressed as:

\begin{equation}
     P(Y_t \in C(X_t)) \geq 1 - \alpha ,
\end{equation}

where \( \alpha \in (0,1) \) is the user-defined error rate. This coverage guarantee is achieved by leveraging a held-out calibration dataset \( \mathcal{D}_{\text{cal}} \) to compute quantile-based thresholds for prediction set construction. The process of applying CP consists of the following steps:

\begin{enumerate}
    \item Heuristic Uncertainty Definition: A notion of uncertainty is defined for the model.
    
    \item Conformal Score Computation: A score function \( s(X_t, Y_t) \in \mathbb{R} \) is defined to quantify the uncertainty of a predicted label \( Y_t \in \mathcal{Y} \) with respect to the input \( X_t \). 

     \item Threshold Calculation: The calibration dataset is used to compute a threshold $\hat{q}$ based on the conformal scores:

    \begin{equation}
        \hat{q} = \text{Quant} \left( \{ s_i : (X_i, Y_i) \in \mathcal{D}_{\text{cal}} \}, \frac{\lceil (n+1)(1-\alpha) \rceil}{n} \right),
    \end{equation}

    where \text{Quant} represents the quantile function used to compute the (1-$\alpha$) quantile of conformal scores in the calibration set, \( s_i \) represents the conformal score for each calibration point, and \( n \) is the number of observations in the calibration dataset \( \mathcal{D}_{\text{cal}} \).

    \item Prediction Set Construction: Given a test input \( X_t \), the prediction set is generated as:

    \begin{equation}
        C(X_t) = \{ Y \in \mathcal{Y} : s(X_t, Y_t) \leq \hat{q} \}.
    \end{equation}

\end{enumerate}

\subsubsection{Conformal Score Functions}
In this work, we consider three conformal score functions: \textit{Least Ambiguous Classifier (LAC)} \cite{sadinle2019least}, \textit{Adaptive Prediction Sets (APS)} \cite{romano2020classification}, and \textit{Hybrid Spatial Score (HSS)}, which map the TSR and OCR confidence scores to a statistically notion of uncertainty.

\begin{itemize}
    \item \textbf{LAC Score Function}: This score function captures the least reliable component by taking the minimum of the TSR and OCR confidence scores. Formally:

    \begin{equation}
        s_{\text{LAC}}(X, Y) = 1 - \min\left(\text{conf}_{\text{TSR}}(X, Y), \text{conf}_{\text{OCR}}(X, Y)\right),
    \end{equation}

    where $\text{conf}_{\text{TSR}}(X, Y)$ and $\text{conf}_{\text{OCR}}(X, Y)$ represent the confidence scores from TSR and OCR for the predicted cell \( Y \) given an input table cell \( X \).
    LAC assigns higher uncertainty scores when either TSR or OCR has low confidence, suggesting that predictions with unreliable structural or textual information receive a higher uncertainty estimate.

    \item \textbf{APS Score Function}: The APS method sums the confidence scores of TSR and OCR predictions to generate the overall reliability of cell predictions:

    \begin{equation}
        s_{\text{APS}}(X, Y) = \sum_{Y' \in \mathcal{Y} : \text{conf}(X, Y') \geq \text{conf}(X, Y)} \text{conf}(X, Y'),
    \end{equation}

    where \( \text{conf}(X, Y) \) represents the confidence score for label \( Y \) given input \( X \). Unlike LAC, APS represents the cumulative confidence of the predictions, addressing cases where individual confidence scores may misrepresent the prediction's overall reliability.

    \item{\bf HSS Score Function}
    Although LAC and APS provide useful uncertainty measures, they may not fully capture the spatial and structural dependencies inherent in table data. Therefore, we first compute two scores, which measure structural and content reliability by integrating row, column, and text confidences derived from the TSR and OCR models. Specifically, we define the \textbf{structural reliability score} based on TATR-derived row and column confidence scores, and the \textbf{content reliability score} using OCR-derived text confidence. 
    
    \paragraph{Structural Reliability:} Given the row confidence \( \text{conf}_{\text{row}} \) and column confidence \( \text{conf}_{\text{col}} \) obtained from TATR, the structural reliability score \( R_{\text{struct}} \) is computed as:
    
    \begin{equation}
        R_{\text{struct}} = \sqrt{(1 - w_{\text{row}} (1 - \text{conf}_{\text{row}})) \cdot (1 - w_{\text{col}} (1 - \text{conf}_{\text{col}}))},
    \end{equation}
    
    where \( w_{\text{row}} \) and \( w_{\text{col}} \) are row and column weight parameters between 0 and 1 that control the influence of TSR confidence scores.
    
    \paragraph{Content Reliability:} The OCR confidence score \( \text{conf}_{\text{ocr}} \), representing the probability that the extracted text is correct, contributes to the content reliability score \( R_{\text{content}} \), given by:
    
    \begin{equation}
        R_{\text{content}} = w_{\text{text}} \cdot \text{conf}_{\text{ocr}},
    \end{equation}
    
    where \( w_{\text{text}} \) is a weight parameter balancing the contribution of OCR confidence.
    
    The conformal score \( s_{\text{HSS}}(X, Y) \) is computed by the geometric mean of the structural and content reliability scores:
    
    \begin{equation}
        s_{\text{HSS}}(X, Y) = 1 - \sqrt{R_{\text{struct}} \cdot R_{\text{content}}}.
    \end{equation}

\end{itemize}

To optimize these weight parameters (\( w_{\text{row}}, w_{\text{col}}, w_{\text{text}} \)), we conduct a grid search to find the best combination that minimizes the prediction set size while maintaining a target coverage level. Through this optimization, HSS aims to adapt to varying table structures and improve uncertainty quantification in scenarios where OCR predictions may be incomplete or unreliable.

In addition to comparing LAC, APS, and HSS, we also performed ablation studies in which the confidence scores from TSR and OCR are treated as the conformal scores without aggregation. This comparison assesses their effectiveness in accurately flagging incorrectly extracted table cell data.

\subsection{Calibration and Uncertainty Score}
The calibration step calculates thresholds denoted as \( \hat{q}_{\text{LAC}} \) for LAC and \( \hat{q}_{\text{APS}} \) for APS using the defined score functions on the held-out calibration data (a randomly selected 50\% subset of the extracted cells from each domain). During inference, we generate prediction sets by comparing the conformal scores for TSR-OCR predictions against these thresholds. To estimate the uncertainty score for a prediction set, we compute:

\begin{equation}
    U(X) = \max(0, s(X) - \hat{q}),
\end{equation}

where \( U(X) \) represents the uncertainty score, \( s(X) \) is the conformal score computed using LAC, APS, or HSS, and \( \hat{q} \) is the threshold obtained from the calibration dataset. A higher uncertainty score indicates a greater likelihood of extraction errors, which are then flagged by a thresholded labeler for human verification.

\section{Experiments}
\subsection{Conformal Score Function Selection}
Selecting an appropriate conformal score function is crucial for effective CP, as it determines how well the model can distinguish between correctly and incorrectly extracted table cells. An ideal score function should provide a well-calibrated measure of uncertainty that correlates with extraction correctness, ensuring that incorrect extractions receive higher uncertainty scores and are appropriately flagged.

To evaluate the effectiveness of different score functions, we tested a range of thresholds (0.01 - 1) for each function and identified the optimal threshold (0.03) for APS score function that maximized the F1-score for our dataset. The F1-score is chosen as the evaluation metric because it balances precision (the proportion of flagged cell contents that are truly incorrect) and recall (the proportion of the actual incorrect cell contents that are successfully flagged). For each score function, we computed precision, recall, and F1-scores across a series of thresholds, selecting the best threshold for each function that yielded the highest F1-score.

Table~\ref{tab:score_fns} presents the mean F1-scores achieved by various score functions. Among the tested functions, the APS score function consistently outperforms others, demonstrating its effectiveness in identifying incorrect extractions while maintaining a balanced trade-off between precision and recall. Based on this result, we select APS as the primary conformal score function for further evaluation of our TSR-OCR-UQ framework.

\begin{table}[h]
    \centering
    \caption{A comparison of average F1-scores of conformal score functions considered.}
    \label{tab:score_fns}
    \begin{tabular}{c|c|c|c|c|c}
        \toprule
        \textbf{Score Function} & \textbf{APS} & \textbf{HSS} & \textbf{LAC} & \textbf{OCR Only} & \textbf{TSR Only} \\
        \midrule
        \textbf{Mean F1-score} & \textbf{0.761} & 0.514 & 0.758 & 0.738 & 0.654 \\
        \bottomrule
    \end{tabular}
\end{table}

\subsection{Experimental Setup}
\begin{figure*}[ht]
    \centering
    \includegraphics[width=1.0\textwidth]{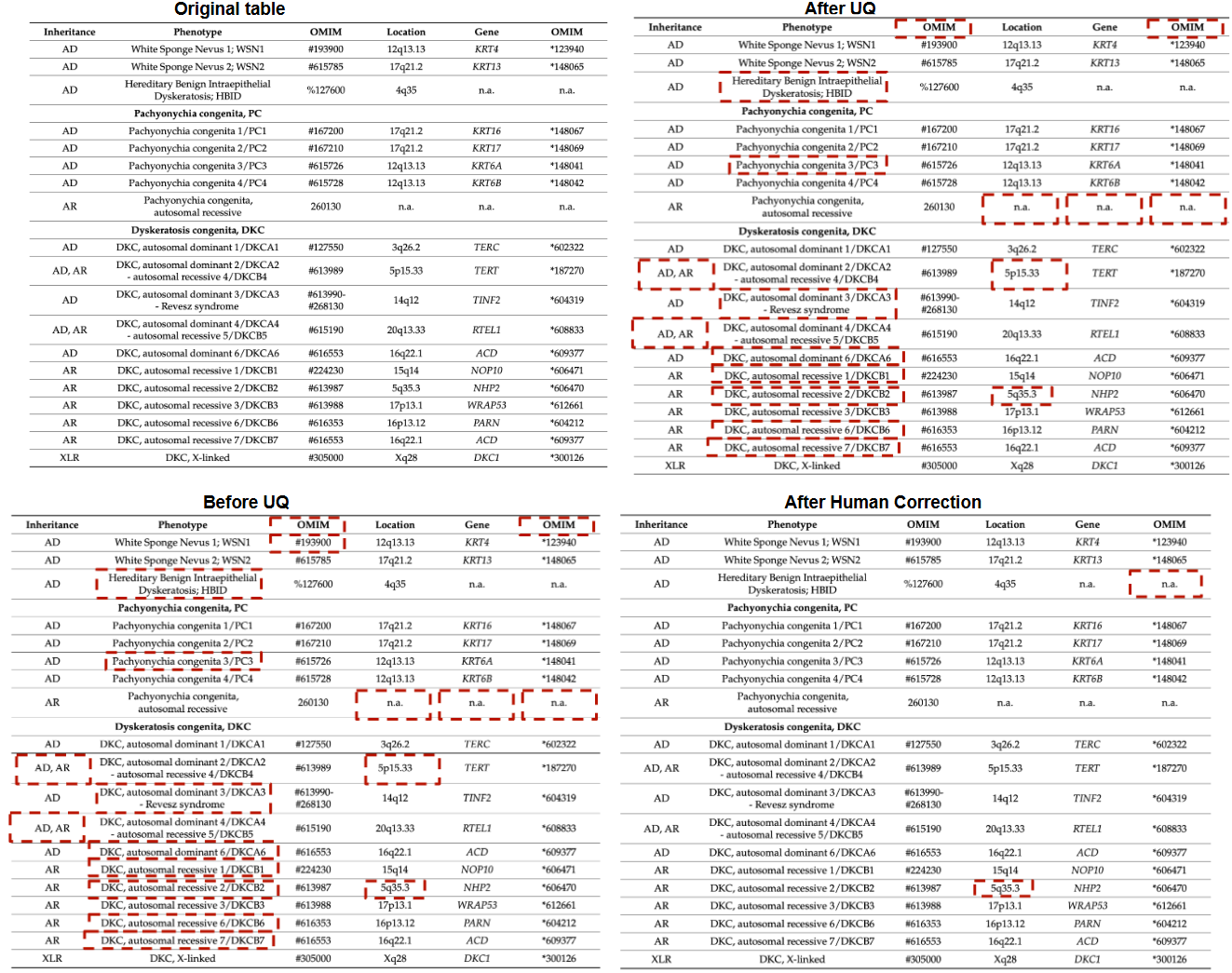}
     \caption{Impact of UQ and human correction emulation on table extraction. Red boxes indicate cells whose coordinates or contents were incorrectly extracted (Before UQ), flagged cells (After UQ), and remaining errors (After Human Correction).}
     \label{fig:tsr_ocr_human}
\end{figure*}

We evaluate our TSR-OCR-UQ framework using complex scientific tables in the GenTSR \cite{ajayi2023study} and ICDAR-2013 \cite{icdar2013competition} datasets. For our experiments, we built a diverse test set comprising 16 tables from Computer Science, 16 from Materials Science, 7 from Biology, and 20 from ICDAR-2013 \cite{icdar2013competition}. Each table undergoes three evaluation phases (see Fig.~\ref{fig:tsr_ocr_human}):

\begin{itemize}
    \item {\bf Before UQ}: This phase evaluates the initial data extraction quality by the framework composed of only TSR+OCR. At this stage, the errors of extracted data are attributed to TSR segmentation, OCR errors, and misalignment between the TSR and OCR results. The extraction performance is measured using data accuracy and error rate, defined as:
    \begin{equation}
        \text{Data Accuracy}_{\text{Before UQ}} = \frac{\text{Number of correctly extracted cells}}{\text{Total number of extracted cells}},
    \end{equation}

    \begin{equation}
        \text{Error Rate}_{\text{Before UQ}} = 1 - \text{Data Accuracy}_{\text{Before UQ}}
    \end{equation}

    \item {\bf After UQ}: At this stage, the system assigns an uncertainty score (Eqn. 10) to each extracted cell. Cells whose uncertainty scores are higher than the threshold are flagged as potentially incorrect. The effectiveness of the UQ framework is evaluated  using the precision and recall, defined as:
    \begin{equation}
        \text{Precision}_{\text{UQ}} = \frac{\text{Number of flagged cells that are actually incorrect}}{\text{Total number of flagged cells}},
    \end{equation}

    \begin{equation}
        \text{Recall}_{\text{UQ}} = \frac{\text{Number of flagged cells that are actually incorrect}}{\text{Total number of actually incorrect cells}}
    \end{equation}
    These metrics assess the framework's capability to correctly identify incorrect extractions.
    In addition, we evaluate the efficiency of UQ in reducing manual verification effort through the labor savings ratio, computed as:

    \begin{equation} \text{Labor Savings} = \frac{N_{\text{all-cells}} - N_{\text{UQ}}}{N_{\text{all-cells}}}, \end{equation}

    where  \( N_{\text{all-cells}}\) represents the total number of extracted table cells, and \( N_{\text{UQ}}\) represents the number of flagged cells requiring human review. A higher labor savings value indicates fewer cells require manual correction, demonstrating the efficiency of UQ in reducing human effort required to verify all the extracted table cells.

    \item {\bf After Human Correction}: This phase evaluates the data extraction quality after human correction of flagged cells. Because we have the ground truth for each cell, we emulate the function of a human to verify and correct cells flagged based on UQ. The remaining extraction errors after this correction are quantified using the error rate:

    \begin{equation}
        \text{Error Rate}_{\text{After-HC}} = \frac{\text{Number of remaining incorrect cells after correction,}}{\text{Total number of extracted cells}}
    \end{equation}
    
    A comparison between the error rate after human correction and before UQ quantifies how well human corrections improve the overall extraction accuracy.
\end{itemize}

\subsection{Experimental Results}
\subsubsection{Baseline Evaluation (Before UQ)}
Fig.~\ref{fig:before_afterHC} presents the data extraction accuracy ($\approx 0.53 - 0.71$) and the error rate ($\approx 0.28 - 0.47$) of the TSR-OCR pipeline before the application of UQ. At this stage, there is no flagging of incorrect extractions; instead, the extracted table data is directly evaluated against the ground truth to assess the overall quality of the extracted data. The results indicate that a substantial number of extracted cells contain errors, which highlight the inherent limitations of a TSR-OCR pipeline and underscore the necessity of incorporating an uncertainty-aware mechanism to improve data quality.

\subsubsection{Effectiveness of UQ on Error Flagging (After UQ)}
Applying UQ using CP with APS score function is effective in detecting data extraction errors, as shown in Table~\ref{tab:after_uq}. Specifically, the overall recall is above 65\%, showing that a significant fraction of incorrect extractions are flagged. Similarly, the overall precision is about 69\%, indicating that most flagged extraction errors are true errors. In addition, UQ improves efficiency, with an overall labor savings of 53\%, meaning that only 47\% of extracted cells require manual review

\begin{table}[h]
\centering
\caption{TSR-OCR-UQ performance in flagging incorrect extractions.}
\begin{tabular}{lccccc}
\hline
\textbf{Domain}  & \textbf{Precision} & \textbf{Recall} & \textbf{F1-score} & \textbf{Labor Savings}\\
\hline
Computer Science & 0.712 & 0.874 & 0.754 & 0.384 \\
Materials Science & 0.540 & 0.738 & 0.565 & 0.605 \\
Biology  & 0.516 & 0.642 & 0.479 & 0.569\\
ICDAR-2013 & 0.481 & 0.553 & 0.563 & 0.563\\
ALL & 0.697 & 0.652 & 0.590 & 0.530 \\ 
\hline
\end{tabular}
\label{tab:after_uq}
\end{table}

These results demonstrate that the UQ in our framework effectively detects most extraction errors and reduce human verification effort, but there is still room to improve the detection accuracy by removing false positives and false negatives. 

\subsubsection{Effectiveness of UQ-informed Human Correction}
\begin{figure*}[h]
    \centering
    \includegraphics[width=1.0\textwidth]{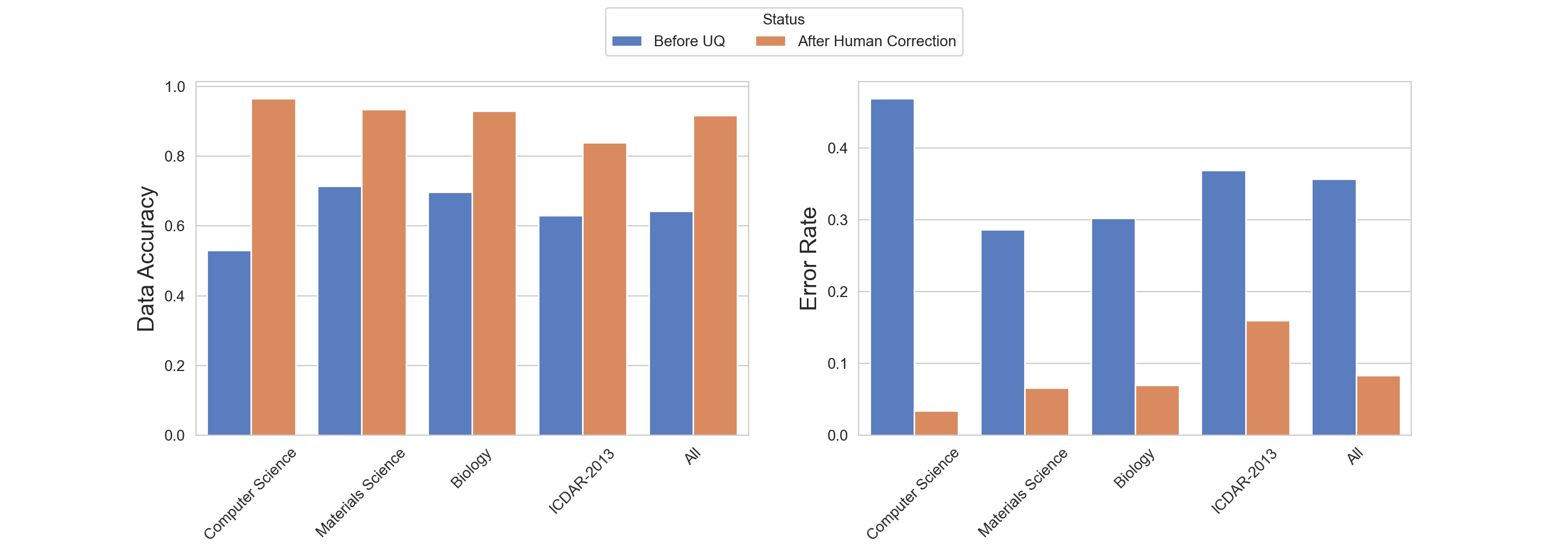}
     \caption{Data accuracy (left)and Error rate (right) before UQ and after human correction across tables in 4 domains.}
     \label{fig:before_afterHC}
\end{figure*}

The results in Fig.~\ref{fig:before_afterHC} indicate that UQ-informed human intervention successfully eliminates the majority of extraction errors without requiring a full review of all extracted table cells. The accuracy of the extracted data improves by about 30\%, increasing from 53\% to 83\% overall. In particular, the data accuracy of Computer Science is improved the most,  from 53\% to 97\%.

\section{Discussion}
Our study demonstrates that incorporating UQ into TSR-OCR pipelines substantially improves table data extraction quality by flagging errors and enabling UQ-informed human correction. However, several observations warrant further discussions.

One key finding is that the baseline TSR-OCR pipeline exhibits high error rate across all disciplines, indicating that many incorrect extractions are present in the results of state-of-the-art TSR-OCR tools. Although applying UQ improves precision, a fraction of correctly extracted cells are mistakenly flagged as incorrect. Our analysis reveals that these falsely flagged cells often exhibit characteristics such as short text strings (e.g., single-character values like ``O'' or ``5''), low OCR confidence despite correct extraction, or alignment ambiguities where text appears near table boundaries. These cases suggest that OCR confidence alone may not always be a reliable indicator of correctness, especially for sparse or ambiguous content.

Additionally, the effectiveness of UQ depends on the choice of the conformal score function. The APS score function outperforms the other scoring functions considered in this paper. However, certain table structures—such as those with multi-row or multi-column spanning cells—exhibit different error patterns that may require adaptive uncertainty quantification. For example, our framework performs best in aligning TSR and OCR results in tables with both well-defined row-column structures and nested tables, but its effectiveness reduces when handling tables with long content in their rows and specialized symbols, many of which are found in Biology and Materials Science.

The differential effort of human correction across various domains underscores the importance of an interactive system. Scientific tables with highly structured data (e.g., in Computer Science) require less human correction compared to domains with noisier OCR outputs, such as Materials Science, where complex symbols and equations can lead to more extraction errors. A potential solution is to explore adaptive human-in-the-loop strategies where correction priorities are dynamically assigned based on the confidence distribution across table regions.

\section{Conclusion}
In this study, we proposed an uncertainty-aware TSR-OCR-UQ framework that integrates TSR, OCR, and Conformal Prediction-based UQ to improve table data extraction quality. Our results demonstrated that applying UQ reduces the potential effort of post-extraction verification and correction by approximately 53\%. We compared various conformal scoring functions and found that Adaptive Prediction Sets outperforms the others. Our findings highlight the effectiveness of UQ in improving table extraction quality but also reveal special cases where UQ is less effective, including ambiguous text, spanning cells, and domain-specific OCR limitations. Addressing these challenges will require further refinements in uncertainty modeling, hybrid automation-human workflows, and more adaptive scoring methods tailored to the diversity of scientific table structures.

\section{Acknowledgement}
The work done by Yi, He has been supported in part by the National Science Foundation (NSF) under Grant Nos. IIS-2236578, IIS-2441449, IOS-2446522, and the Commonwealth Cyber Initiative (CCI).

\bibliographystyle{unsrt}
\bibliography{references}

\end{document}